\documentclass[submission,copyright]{eptcs}
\providecommand{\event}{--} 
\usepackage{breakurl}             
\usepackage{graphicx}
\usepackage{amssymb}
\usepackage{amsmath}

\title{Analysis of Software Binaries for Reengineering-Driven Product Line Architecture -- An Industrial Case Study}

\author{
Ian D. Peake, Jan Olaf Blech, Lasith Fernando
\institute{RMIT University, Melbourne, Australia}
\and
Divyasheel Sharma, Srini Ramaswamy, Mallikarjun Kande
\institute{ABB Corporate Research, Bangalore, India}
}

\def\titlerunning{Reengineering-Driven Product Line Architectures}
\def\authorrunning{Peake et al.}
\begin{document}
\maketitle

\begin{abstract}
This paper describes a method for the recovering of software architectures from a set of similar (but unrelated) software products in binary form.  One intention is to drive refactoring into software product lines and combine architecture recovery with run time binary analysis and existing clustering methods. Using our runtime binary analysis, we create graphs that capture the dependencies between different software parts. These are clustered into smaller component graphs, that group software parts with high interactions into larger entities. The component graphs serve as a basis for further software product line work. In this paper, we concentrate on the analysis part of the method and the graph clustering. We apply the graph clustering method to a real application in the context of automation / robot configuration software tools.
\end{abstract}

\section{Introduction}

In large organizations, due to mergers and acquisitions over multiple years, sometimes decades apart, multiple products evolve and mature to serve different market segments. These may have similar functionalities derived from similar market and product requirements. Although these are candidates for software product line approaches, they differ in their architectural evolution, due to factors such as technology lock-ins, customer driven technology choices, talent availability and pipeline and outsourcing silos.
Product lines approaches have been shown to significantly reduce costs and improve productivity in certain industrial contexts, and have therefore been considered attractive.
However there are relatively few corresponding experience reports in the literature. In particular in contexts as described above, it is not surprising if product line engineering must focus on so-called ``reengineering-driven'' scenarios, where any planned reuse must first account for a set of products with a complex mix of common features, subtle semantic differences and little implementation overlap.
In such large organizational contexts it is challenging to discuss possible reuse across products  because of the difficulty of sharing or discussing development assets such as requirements or architecture documentation or source code, let alone the discussion or generation of ``big picture'' knowledge relating architectures between products.

This paper describes a method and a case study seeking to derive structural software architecture descriptions of unrelated but similar software products, from software binaries, into a form intended for evaluating and driving refactoring into software product lines.
Our industry partner ABB has several organizationally separate multi-national divisions with partly distinct business goals and accountabilities, where each division provides one or two products which exhibit similar requirements and functionality.
We investigate a software product line architecture scenario in ABB involving a set of software products for control automation engineering.
However for these products it is challenging both to elicit requirements which propagate from business units to tools and to obtain information about existing tool architectures (tracing requirements to implementations). Efforts in this exercise focus on extracting software architecture documentation from documentation and binaries, with the goal to use feedback from draft documentation to propose a software product line architecture. For this purpose we focus on an architectural description notation based on graphs which formalizes dependence relationships between components at arbitrary levels of abstraction. Use of graphs has the advantage that they are suitable for use as input/output to and from tools as well as intuitively understood by humans.

For this work, we develop a method for architecture recovery based on runtime and static binary analysis use it to generate documentation. Typically for scenarios involving architecture recovery, the scale of graph recovered is is a major challenge. Even a relatively coarse-grained analysis of most systems provides too much detail to be useful as documentation. Also it is desirable to relate candidate components to functionality. We use an existing clustering method to try to resolve these issues. We examine binaries using Intel's Pin tool~\cite{pin} for binary instrumentation, as well as free and open source tools for static analysis of generic MS Windows executables and .NET assemblies.
The contribution of this paper is:
\begin{itemize}
\item
  	Our method: a description of the architecture derivation process and its adaptation to the systems of interest and the realization of the graph clustering in this context.
\item
	A case study in deriving structural software architecture descriptions from software binaries intended for engineering-driven software product line architecture.
\item
	An evaluation of the method and lessons learned.
\end{itemize}
A preliminary 6-pages work in progress paper has been published \cite{peake}, featuring parts of the method but no application to an industrial case study.

\section{Related Work}

\label{sec:relwork}

Although software product line architecture is shown to improve productivity significantly in certain industrial contexts, there are relatively few methods and experience reports in the literature related to reengineering-driven scenarios. Software product line engineering process models typically assume either (i) a centralized and top-down management-driven initiative to introduce and support reuse efforts or (ii) an initial single effort at source code level to develop a product which has splintered in an ad hoc way through code forking into multiple related projects. Moreover in the largest organisations the existence of independent product groups makes it challenging to meaningfully share or even to discuss, at a technical level, sharing development assets such as architecture documentation and source code.

Reengineering-driven product line architecture minimally requires architecture documentation for all related products, however few existing (or recent) efforts focused specifically on methods for architecture reconstruction for software product lines.
Fewer still focus on methods explicitly suited to the reengineering scenario.
The MAP method---``mining architectures for product lines''~\cite{Stoermer2001}---assumes shared knowledge of product lines at management and technical level and availability of people, in particular architects, source code, documentation and thus emphasis maximizing stakeholder buy-in, need for access to software architects and
developers.
Another assumption is that candidate products are all initially situated in similar domains.
The MAP method consists of several phases: preparation, extraction, composition,
qualification, evaluation and follow-on activities.
The extraction phase for each product is
concerned with extraction of an implementation model from existing source assets (note
assumption that source code is available), and jointly, abstraction to an component-based
architectural model and mapping known architectural styles and attributes
onto the architectural model.
Other related work~\cite{Pinzger2004} described a method and experience in architecture recovery for product families. Notably it is argued ``Apparently, the process step of abstracting meaningful higher-level views from the low-level model or source code is the most complex task. A lack of appropriate (semi-)automatic tools increases manual work.'' Our approach provides a solution for one step in an automated tool based approach.

Breivold~\cite{Breivold2008} and Kettu~\cite{Kettu2008} describe case studies in analysis of industrial automation
systems in ABB. At least one author is common to these publications.
Breivold describes a case study in analysis of an industrial automation control system
to improve its evolvability using their architecture evolvability analysis (AREA) method. The
method is not specifically focused on software product scenarios.
Kettu describes experience and lessons learned, synthesised as practical advice,
and illustrated by two case studies in architecture reconstruction of ABB industrial
automation systems. There is mention of a system called DependencyTool, developed for
the static analysis of binaries, to overcome difficulties in analysis of source code such as C
which typically requires preprocessing before analysis.
Another case study describes the development of a PLA for ABB Robotics software~\cite{koziolek2009}.
A more recent industrial case study on software product lines in the industrial automation domain has been conducted~\cite{koziolek2014}. From a methodological view, this is complementing our runtime analysis part.

Existing published work on architecture reconstruction and related reverse
engineering tasks focussing on derivation of component candidates and inter-dependencies is covered in existing surveys and overview papers \cite{ChikofskyandCross1990,Pollet2007,Canfora2011}.
Two main directions are highlighted: 1) methods based on analysis of source code and 2) methods
based on the analyses or execution of system binaries.
A taxonomy of reverse engineering techniques includes classifications according to the artefacts used, and whether analysis is static---based on syntactic analysis of source or executable---or dynamic---based on running, observing or animating the system itself~\cite{Canfora2011}.
In this paper,  we focus on runtime analysis for architecture derivation. This is also called dynamic analysis~\cite{Cornelissen2009}.
DiscoTect \cite{yan2004,Schmerl2006} is an existing framework that reconstructs architectures of running systems, designed to handle multiple high level architectural styles and possible realizations in implementations.  Its DiscoSTEP language enables mappings to be manually defined by domain experts for interpreting low level system events as more abstract architectural elements defined as coloured Petri Nets. DiscoTect analyses execution traces collected by a trace engine like method calls between objects using the Java Platform Debugger Architecture.

Reconstructing software architecture from execution traces is structured into two tasks 1) the analysis of the execution traces and 2) the identification of potential components. Combining potential {\em component candidates} into disjunct sets denoting suggestions for aggregation of components is known as clustering. It is an important step for gaining suggestions on the original and potential future architectures. The field of clustering for software components has been studied by several authors including \cite{Lung1998}. In \cite{Mitchell2001} an analysis of source code for component detection is featured.  \cite{Koschke2002} studies clustering in the context of software evolution.

In this work we are using the Pin tool \cite{pin} for binary
instrumentation and tracing hints about the architecture at runtime. Other well known
tools include the Valgrind~\cite{valgrind} tool which does
not have native Windows support. It offers, however, a wider range of
instrumentation possibilities potentially resulting in slower code.

\section{Case Study}

An overview of the candidate systems under study is given in Figure~\ref{fig:overview}.
For confidentiality reasons no ABB products, subcomponents or other elements can be named.
We list the primary implementation language, size metric and number of installed binary components and comments related to composition.
In many respects systems show some expected diversity in relation to implementation language, style and scale.
All systems are native Windows 32 bit applications.
Binaries for the systems range in size from 32 to 438 megabytes.

\begin{figure}
\centering
{\small
\begin{tabular}{l | lll}
Name & Primary Implementation Language & Size (Mb) / Size (\#binaries) & Composition \\
\hline
S1 &	C++ &	32 / 80 &	Main EXE +  DLLS, COM \\
S2 &	C++ &	133 / 289 &	\\
S3 &	C++ &	196 / 515 &	Main EXE + other EXE \\
S4 &	.NET &	281 / 348 &	\\
S5 &	.NET &	438 / 340 &	\\
S6 &	Other &	41 / 14	& \\
S7 &	C++ &	76 / 165	& Main EXE 
\end{tabular} }
\caption{System overview}
\label{fig:overview}
\end{figure}

In Figure~\ref{fig:reuse} we attempt to quantify which components from each existing system are reused or shared by any other, using a simple metric.
It seems plausible that one or more components in binaries may be reused or shared across products in the organization, with consistent names.
For a given pair of systems, our metric counts the number of filenames which are common to both systems' install directories.
For each system we take a recursive directory listing of the (binary) install directory and create a list of files in the directory and canonicalise them by translating characters to lower case.
Then we count the number of filenames which appear in both systems' listings.
The measure is symmetric.
Only non-zero entries are shown (where there are some shared files).
The measure roughly matches our understanding of the systems:
S1 and S2 are closely related, with significant implementation overlap, while the pairs (S2,S7) and (S3,S6) share a single common component but are otherwise unrelated. The metric is approximate for several reasons: first, filenames do not reflect the binary contents, i.e. filenames could be renamed without affecting the behaviour of components.

\begin{figure}
\centering
{\small
\begin{tabular}{|l|l|l|l|l|l|l|l|}
\hline
Systems & S1 & S2  & S3  & S4  & S5  & S6  & S7  \\
\hline
S1 	& 67 & 53  &     &     &     &     &     \\
\hline
S2 	& 53 & 257 &     &     &     &     & 1   \\
\hline
S3 	&    &     & 509 &     &     & 1   &     \\
\hline
S4 	&    &     &     & 640 &     &     &     \\
\hline
S5 	&    &     &     &     & 680 &     &     \\
\hline
S6 	&    &     &  1  &     &     & 14  &     \\
\hline
S7 	&    &  1  &     &     &     &     & 164  \\
\hline
\end{tabular} }
\caption{Implementation overlap}
\label{fig:reuse}
\end{figure}

Of these systems, binaries for S1, S3, S4, S5 and S7 were selected for analysis based on assessment of analysis feasibility.

\section{Approach}

\subsection{Method}

Our aim is the recovery of architectural constructs as defined by authors such as Szypersky~\cite{Szypersky} and Kazman~\cite{kazman},
focusing on the following, in rough order of priority:
system components---that is, independently deployable units. (We treat deployability as a proxy for reusability);
dependencies on external systems;
relationships between components (and external systems, or their components);
relationships between (user-observable) features and components;
contracts which govern interactions between components;
the most important aspects (functional/extra-functional properties, not only user-visible);
relationships between aspects and components.
The steps in our method are:
\begin{enumerate}
\item {\em System identification}: Identify the candidate software systems for product line refactoring.
\item {\em Requirements definition}: Key requirements in the form of features, extra-functional properties and use cases should be identified.
\item {\em Static and dynamic binary analysis and graph construction}. As discussed, in this step program data in the form of a (detailed) graph.
\item {\em Abstraction}: Form clusters around lower level components which correlate with hypothesised components.
\item {\em Visualization}. Since our method involves several steps which can be automated, in particular abstraction, yet nevertheless are imperfect, human intervention is useful to assess and adjust the method. Visualization provides a human-readable representation of abstract graphs for this purpose.
\item {\em Software product line architecture}: Identify a product line architecture and justify its alignment with the respective architectures and requirements of each software system.
\item {\em Refactoring}: Restructure each of the respective systems to align with the product line architecture.
\end{enumerate}
The emphasis in this paper is on steps 3--5.
Our experience with steps 1--6 applied to the case study system is summarized as follows.
Step 1 is presented in the case study section above and are not otherwise a focus of this paper.
Step 2 is not a focus of this paper---for an example of use cases see~\cite{peake}.
Steps 3--5 and our experience with them are summarized in detail below.
Readers are referred to~\cite{koziolek2014} for more details of methods related to step 6.

\subsection{Static Binary Analysis}

We use both static and dynamic analysis of software binaries to derive dependency graphs as a basis for informal comparison between systems.
We produce for each system a graph where nodes are interpreted as (low level) components and edges are interpreted as directed dependencies with labels identifying caller and callee sites (method names or addresses) within the component. Thus each edge encodes the occurrence of one method from one component calling another method in another component. 

Static analyses are derived from binary metadata. Every system studied is Windows-based and comes installed as a collection of executables (EXE and DLL files) and miscellaneous (documentation, configuration) files, therefore the most apparent level of reuse and interpretation is at executable file level. Windows executable metadata often reveals some coarse dependence structure between executables needed for dynamic loading and linking.

Given the goal to derive documentation for cross-product comparison, it is implicit that dependency graphs for respective products should be of a roughly similar number of components (at least order of magnitude) to enable comparisons by humans, however it is impossible to find a simple interpretation of ``component'' satisfying this constraint, rather deeper, and dynamic, analysis, using different interpretations for different systems, is needed. For some systems, core functionality is concentrated in one or a few very large executables revealing virtually no structure. To enable informal analysis, we also extract further information, namely packages, classes/objects and/or methods and in particular names, according to implementation language. For most systems, symbol tables are available in individual executables enabled extraction of type and member/method names. For .NET systems we analyse at the level of assemblies. For .NET systems, package, type and member names is present in the executables.

\subsection{Dynamic Binary Analysis}

To clarify architecture information derived statically, we also trace control flow of executables at runtime. We instrument selected systems using Intel’s Pin tool~\cite{pin} which uses just-in-time (JIT) interpretation/compilation to provide functionally non-invasive custom logging, tracing or profiling. We adapt an existing Pin plugin to instrument all direct transfers of control visible in the executable including in imported code, as well as all indirect or dynamically discovered transfers of control. Every (first) occurrence of a transfer of control between a unique pair of source and target addresses is logged in the form of a time stamp, and a source and target object (typically executable) and address. The output of such a process is used to construct a graph.
Static analysis may be a coarse over-approximation of behavior and weak in identifying likely paths, while dynamic (e.g. runtime) analysis suffers from covering only a few use cases. Thus we expected that structure revealed by dynamic analyses might be weak. However, with a surprisingly small amount of instrumentation, it appears that a reasonable amount of data was available about runtime interdependencies between components.
Although we use the terminology of method calling, strictly speaking at trace level we are logging branches and calls and treating these as ``call'' edges. Although strictly we start with source/target addresses at trace level, wherever possible we reconstruct ``called'' method names by inference from symbol tables.

As a refinement of dynamic analysis, we also trace key functionalities or aspects of a tool. We log runtime paths based on specific usage scenarios. The idea is to invoke a distinct function or aspect of a tool using sequences of user interactions with the tool. The component interactions are then extracted from the generated control-flow trace in order to gain hints on architectural details.
We use a small Pin tool which attaches to an already-running process, logs loading of all executables and logs the first occurrence of a control edge along with a ``time stamp'' (counter). The use of time stamps makes it feasible to reconstruct the order of control edge discovery. The ability to start logging midway through a session allows the ability to approximately isolate the correlation between a specific user action and the code triggered.

\subsection{Abstraction}

To generate component dependence diagrams we perform clustering using the LIMBO algorithm~\cite{limbo} to reduce detail and highlight the most-relevant abstractions.

Architecture analysis artefacts such as component dependence diagrams or call graphs as described above may be displayed graphically as a common form of documentation (or indeed as a common form of high level design notation). However in practice the scale of such graphs as generated may make human comprehension of these formats prohibitive. With the engineering tool systems encountered this proved to be the case for several of the tools which consisted of one hundred or more executables.

There is scope for machine abstraction of these graphs using clustering methods. Clustering is a generic method where a collection of objects are partitioned according to their similarity. {\em Objects} are characterized according to a set of multi-valued or real-valued {\em attributes}. Then the basis for similarity between two objects is the extent to which they have similar attributes. Clustering of graphical software views is already a well-studied area with one main objective being to provide useful abstractions for documentation purposes.

We use a clustering algorithm known as LIMBO~\cite{limbo}.
LIMBO is based on a generic information-theoretic method called Agglomerative Information Bottleneck (AIB).
LIMBO has been used for the analysis of large systems across scientific disciplines including for architecture reconstruction.
LIMBO remains one of the known-best approaches for architecture reconstruction,
based on studies comparing results from algorithms and experts on real large systems~\cite{Maqbool}.
LIMBO and the underlying AIB method are generic---they operate fundamentally on a set of objects $O$, a set of attributes $A$ and relation $R \subseteq O\times A$ with non-negative real number weighting $w: O\times A \rightarrow \mathbb{R}^+ \cup \bot$.
A major purpose of attributes is to encode the graph-theoretic dependencies among objects, thus $A \supseteq O$, and a relationship between $o,o' \in O$ is encoded as $w(o,o')\neq \bot$.

Our method uses LIMBO as follows.
Each primitive (component) is modelled both by an object in $o\in O$ and an attribute in $a\in A$.
For a component $o$, $w(o,a)$ is the number of different ways $o$ calls a different object $a$.
$R$ and $w$ are constructed from the execution traces in an application-dependent way.

First, weights are modified via a suitable weighting transformation.
We select term-document frequency weighting in our case, which transforms weights according to their significance (the more rarely held an attribute $A$ is overall by all objects,
and the more frequently by some given object $O$,
the more significant, thus heavily weighted, $A$ is for $O$.)
Next, the new weights are converted to probabilities such that the sum of all weights per object is 1.
Finally, LIMBO attempts to compress its representation of $R$ by iteratively merging the {\em closest} pair of objects and generating a new relation $R'$ which approximates $R$ under merging.
The closest pair is the one for which merging minimises information loss in $R'$.
LIMBO uses an additional phase to analyse the merge history (resembling a tree) to determine an appropriate point in the merge history (hopefully with a small number of objects) as the most appropriate clustering.
LIMBO's genericity enables it to support both ``structural'' and ``non-structural'' attributes.
Structural attributes reflect program dependence structure as described above.
In our work so far clustering is on a structural basis.
Non-structural attributes refer to the general case and cover properties such as a time stamp or authorship.

LIMBO iteratively and greedily searches for and merges the pair of (``similar'') objects which, when merged into a single object approximating both, preserves the most information overall until eventually only a single object is left, recording the merge history. Additional pre- and post-processing phases combine to improve performance for large object sets, and to select the most suitable point late in the merge process to reflect the best clustering. Following a generic recommended method, we take a low-level graph and encode it for clustering.

Functionality tracing is performed repeatedly across all tool sets, to try to informally relate known or hypothesized features / aspects to specific binary artefacts ({\em primitive (component)s}). An obvious method for tracing involves searching based on domain knowledge for components based on the name of a specific feature or aspect, or a synonym, such as ``firmware download'' or ``hardware connection.'' In our case study, several such attempts may be classed as partly successful, however gaining a precise characterization of such mappings, or isolating a single component corresponding to a feature is unsurprisingly difficult. More systematic functionality tracing was performed as followups to other methods below, as a means of informal evaluation.

\subsection{Visualization}

Although our method of generating graphs varies (see above), we aim to apply a common, abstract interpretation on them, based on a notion of dependence between components, where the degree of dependence may vary.
Following a generic method recommended as part of LIMBO, we take a low-level graph and encode it for clustering. The graph format encodes a set of edges, where each edge relates a pair (component1,method1) with another pair (component2,method2). Thus each edge encodes the occurrence of one method from one component calling another method in another component. (Although we use the terminology of method calling, strictly speaking at trace level we are logging control flow branches and calls and treating these as ``call'' edges. Moreover although strictly we start with source/target addresses at trace level, wherever possible we reconstruct “called” method names by inference from symbol tables.
Although the method of generating graphs varies (see above), the interpretation of the graphs is the same in all cases and is based on a notion of dependence between components where the degree of dependence may vary.
We first create a set of objects corresponding to each unique component and set their attributes. The attributes of each component correspond to all low level components. If there is some dependence between C1 and C2, then C1 has the attribute C2, and C2 has the attribute C1 (thus a symmetric interpretation of a dependence graph). The number of unique edges between C1 and C2 determines the value of the attribute C2 for C1 and vice versa. If there is no direct interaction between C1 and C2, the value of C2 for C1 is zero.

\begin{figure}
\centering
\includegraphics[width=0.3\textwidth]{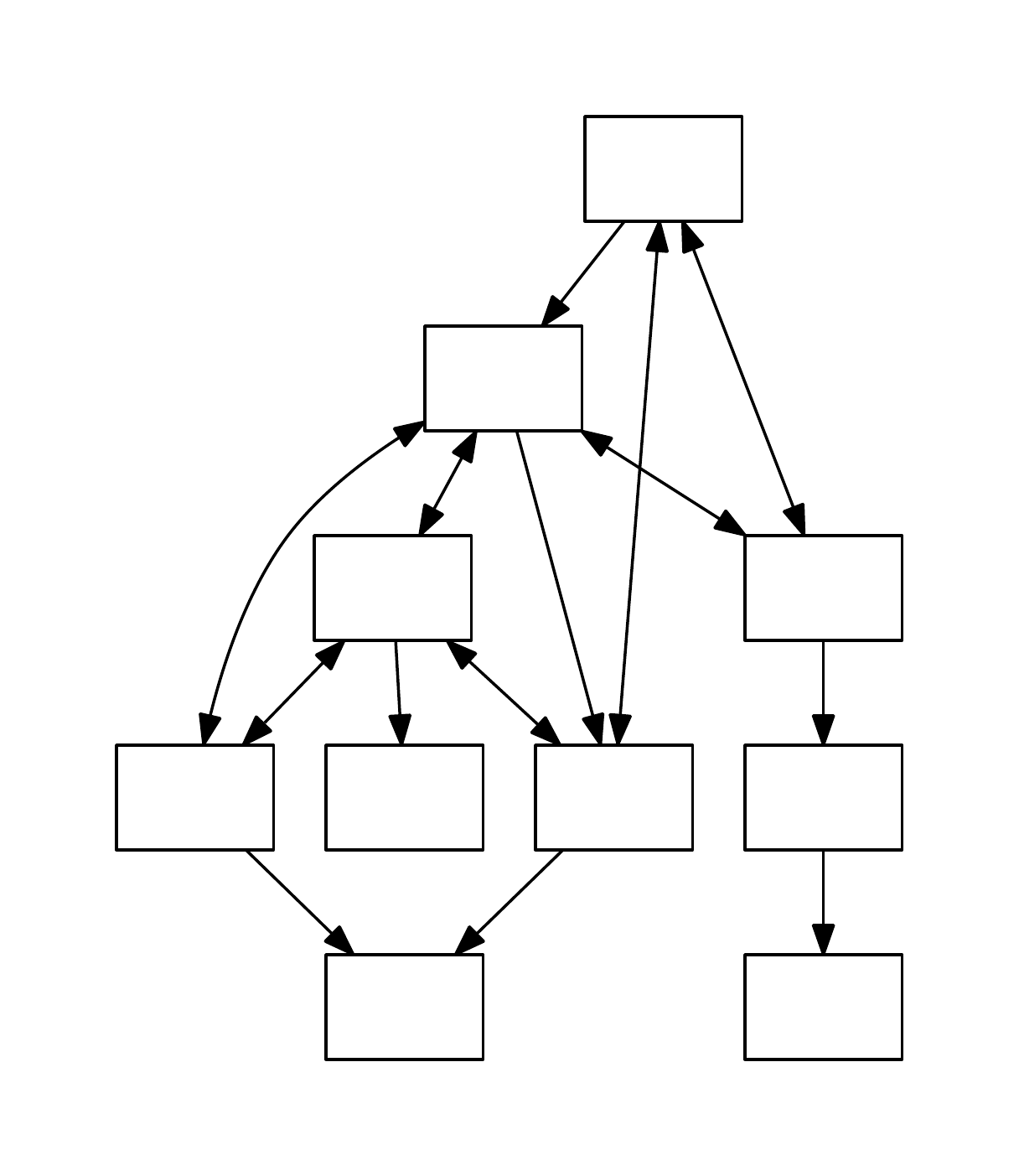}
\caption{S7 as documented}
\label{fig:s7-documented}
\end{figure}

We obtain a clustering of low level components into high level components which we represent graphically (for example Figure~\ref{fig:s7-discovered} below, c.f. the corresponding documented architecture in Figure~\ref{fig:s7-documented}). Nodes are shown as ellipses. For confidentiality reasons no component, object or method names are shown.
The number of unique dependencies between components are provided by labels on the edges. The main dependence direction is given first. Dependencies in the opposite direction are written in brackets. Edges are colour-coded according to how frequently calls/jumps are made between components.
Clustering introduces a second higher level of structure---each cluster may be considered a candidate high level subsystem and requires a name. We settled on a method involving the labelling of clusters with the names of all primitive components of the cluster, with the primitives ranked in descending order by a significance metric. Due to confidentiality reasons an example from the case study cannot be shown. The significance metric favours primitives with more interactions outside of their own cluster, and a higher ratio of outgoing calls to their own cluster to incoming calls from their own cluster. The rationale for such a metric is that: (i) primitives with more overall interactions outside the cluster are de facto ``interfaces,'' hiding significant abstractions, whereas primitives with fewer interactions outside the cluster are more likely internal and incidental; and, (ii) primitives with higher ratio of outgoing calls to incoming calls within the cluster are more likely at a higher level of abstraction.
%
%
\begin{figure}
\centering
\includegraphics[width=0.58\textwidth]{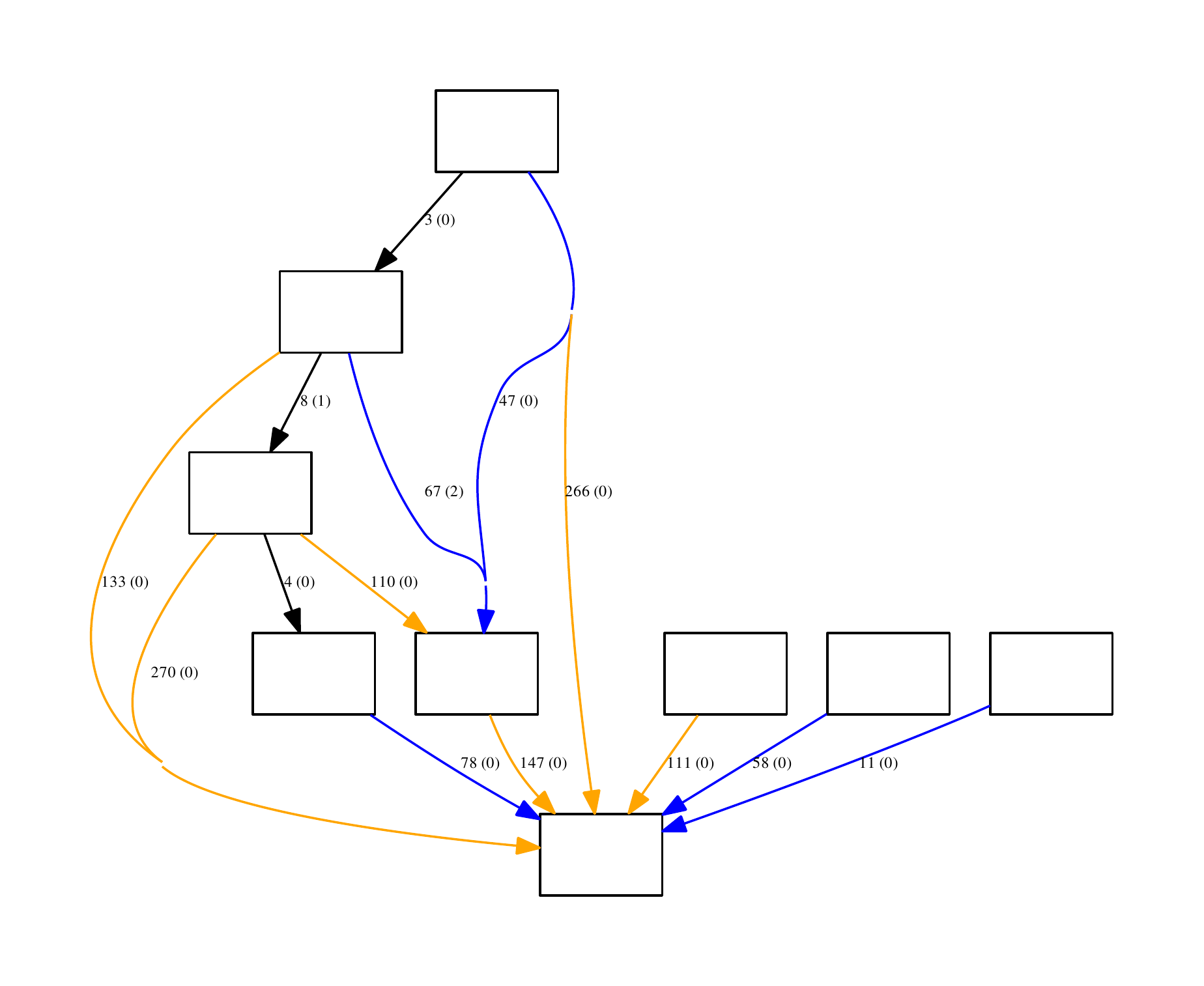}
\caption{S7 - discovered architecture}
\label{fig:s7-discovered}
\end{figure}
\begin{figure}
\centering
\includegraphics[width=0.58\textwidth]{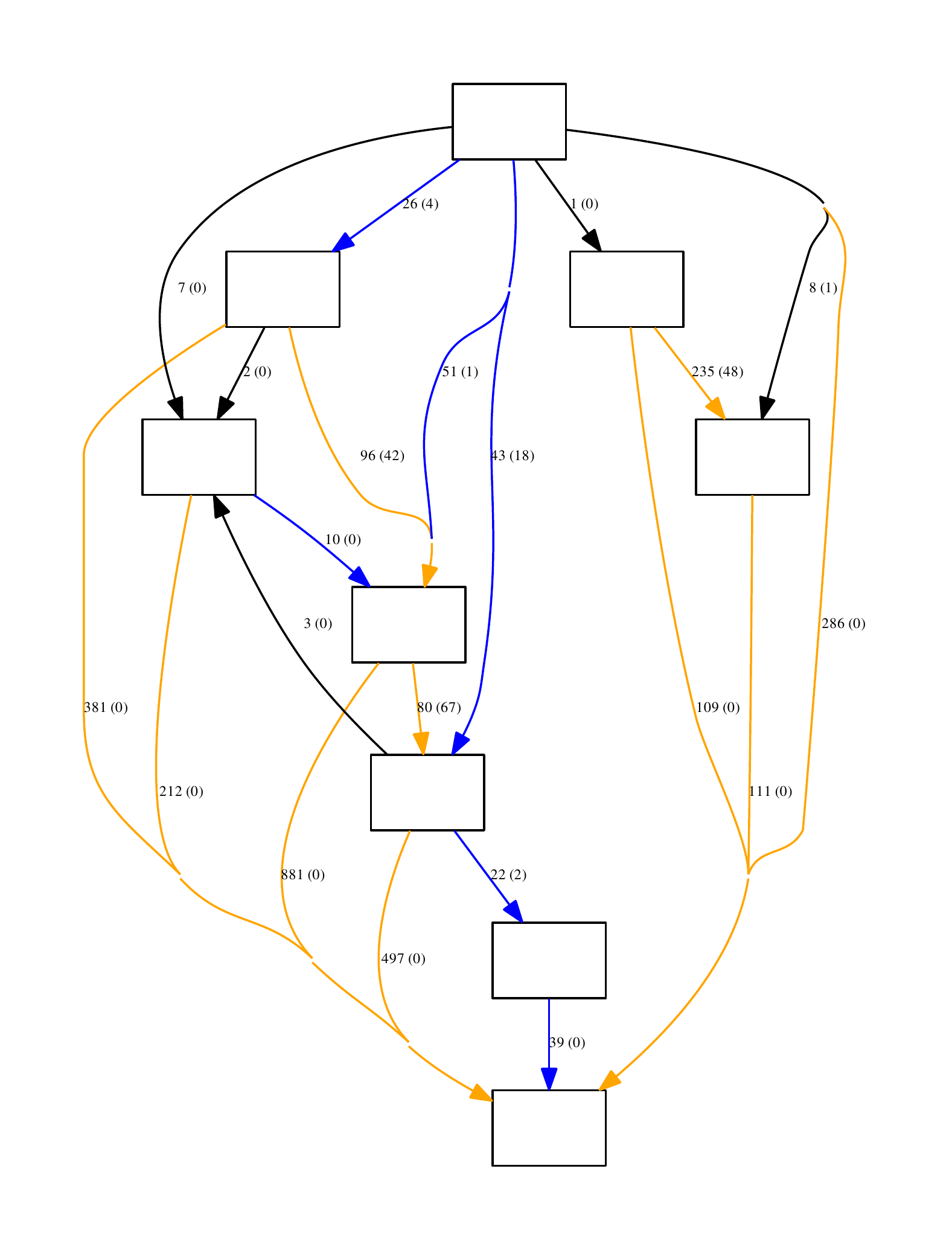}
\caption{S7 - ``refactored'' architecture based on use cases}
\label{fig:s7-refactored}
\end{figure}

\subsection{Software Product Line Analysis}

It is possible to consider a ``refactored'' version of an architecture, based on dynamic information recovered from a selected set of use cases.
For two selected tools (S1 and S7) we recorded and graphed the trace for each of a small set of ``typical'' use cases: load (a project or component), edit (FBD/ST/etc.), check, save, simulate / go online. We extended the clustering method above to provide an ``overlay'' which colours the names of primitive components deterministically, according to a variation of our significance metric above. This work is inspired by, though not systematically following, prior work to trace requirements to code, however it proved useful for validating informal attempts to trace functionality more precisely. In the case of S7 this method appears to have directed analysis to components which might not otherwise have been considered, where names for components (and indeed methods) did not make functionality completely apparent.
In Figure~\ref{fig:s7-refactored} we show an example of the result of a dynamic analysis based on use cases.

\section{Evaluation of Software Binary Clustering}

There are several reasons why clustering may not reflect an authoritative architecture and therefore a number of possible improvements. For example there may be insufficient data in the run time call graph, or architectural anti-patterns may be enforced by the as-is architecture. Where many primitive components are clustered together, it is desirable to associate each component with a meaningful name or feature. Ideally a name is an exact abstraction of the collection of primitive components. This depends on understanding what abstractions (e.g. aspects) are semantically common to all objects of a component, or the principle abstraction of the component. Our ranking metric for the most significant primitives in a cluster could be used to select names from the highest ranked primitives as candidate names for the cluster. Although this has not been evaluated systematically it appears promising in practice.
The clustering method on which LIMBO is based is generic. There is scope to assign so-called ``non structural attributes,'' perhaps based on manual assignment of features, or based on other attributes such as location in a source code hierarchy, perhaps with input from architects or system experts.
Albeit that in this form LIMBO no longer accepts graphs, but rather weighted binary relations.
These could pertain to specific features or aspects such as ``UI'' or ``Safety'' or even identifiers, identifier segments or ``topics'' relating to identifiers or documentation and influenced by domain language as extracted from domain documentation.
For instance we have experimented with some methods described in~\cite{garcia2011}.
Such additional attributes can be assigned to those objects and then taken into account during clustering. There are some existing aspect mining approaches which may be applicable.
The graph data itself could be improved. For runtime analysis, call graphs were generated based on relatively few use cases and short sessions based on introductory tutorials. Thus certain objects are not exercised, have few associated calls, and their corresponding components are probably not being clustered properly.
There is ambiguity in the literature about whether to generate structural attributes by treating the primitive call graph as directed or undirected. Treating the graph as directed results in asymmetric attribution---the relationship between two components is represented in their two respective objects by two separate attributes with distinct values depending on which component is the ``caller''---whereas treating the graph as undirected results in the two objects having attributes with the same value. In our work the call graph is interpreted as undirected, which seems to improve clustering for certain components which call few other components.

Garcia and others introduce the notion of ground truth in software architecture~\cite{garcia2012}, defined as a reliable, authoritative architecture which has been certified correct by a
long-term contributor, that is, someone with long term involvement in a project and intimate
knowledge of the system’s architecture. They propose a framework for
establishing ground truth architectures which minimize the involvement of long-term
contributors, involving a set of processes and a set of so-called mapping principles for
grouping code-level entities into architectural elements and identifying interfaces. Mapping
principles are prioritized according to their classification as either application, generic,
domain principles, in descending order of priority. Such approaches seem likely to be applicable in 
our context.

\section{Conclusion}

In this paper we presented our method for extracting software system architectures into a graph based format that allows further analysis. We presented a case study exemplifying the use for software product line-based work. 
Future work comprises the use of graphs obtained from the method described in this paper as a formal notation for a variety of analysis tasks (such as comparing variants of a product line) and as a basis for more automated product line identification and refactoring. While the graphs are use case based, they can even serve as a representation for formally proving the absence of interdependencies between different components, under a given use case.

\nocite{*}
\bibliographystyle{eptcs}

\end{document}